\documentclass[12pt,a4paper]{article}
\usepackage{cite}
\usepackage{fullpage}
\usepackage[centertags]{amsmath}
\allowdisplaybreaks[4]
\usepackage{amssymb}
\usepackage{bm}
\usepackage{url}
\usepackage[dvips,hyperindex]{hyperref}
\begin{document}

\begin{flushright}
\begin{tabular}{r}
\textsf{7 January 2010}
\\
\texttt{arXiv:1001.0760v2 [hep-ph]}
\end{tabular}
\end{flushright}
\vspace{1cm}
\begin{center}
\Large\bfseries
No Effect of Majorana Phases in Neutrino Oscillations
\\[0.5cm]
\large\normalfont
C. Giunti
\\[0.5cm]
\normalsize\itshape
INFN, Sezione di Torino,
Via P. Giuria 1, I--10125 Torino, Italy
\end{center}
\begin{abstract}
It is shown that the Majorana phases of the neutrino mixing matrix
cannot have any effect in neutrino oscillations,
contrary to the argument presented in \texttt{arXiv:0912.5266}.
It is also shown that
in a charged-current weak interaction process it is not possible to
create a coherent superposition of different flavor neutrinos which is independent of the associated
charged leptons.
\end{abstract}

It has been argued in Ref.~\cite{0912.5266}
that it is possible to observe the effect of the Majorana phases of the neutrino mixing matrix
in a neutrino oscillation experiment with an initial beam described by a superposition
of different neutrino flavors.
Unfortunately, the argument presented in Ref.~\cite{0912.5266} is invalid, as shown in the following.
It is also in contradiction with the well known proof that
the Majorana phases do not
have any effect in neutrino oscillations in vacuum
\cite{Bilenky:1980cx,Doi:1980yb,Schechter:1981gk}
as well as in matter
\cite{Langacker:1986jv}.

In the following we consider the simple
$\nu_{e}$-$\nu_{\mu}$
two-neutrino mixing framework considered in Ref.~\cite{0912.5266}.
In this framework,
the left-handed flavor neutrino fields $\nu_{eL}(x)$ and $\nu_{\mu L}(x)$
are unitary combinations of the left-handed massive neutrino fields
$\nu_{1L}(x)$ and $\nu_{2L}(x)$ with respective masses $m_{1}$ and $m_{2}$
(see Ref.~\cite{Giunti-Kim-2007}):
\begin{equation}
\nu_{aL}(x)
=
\sum_{k=1,2} U_{ak} \nu_{kL}(x)
\qquad
(a=e,\mu)
\,,
\label{001}
\end{equation}
where $U$ is the $2\times2$ unitary mixing matrix.
Flavor neutrinos are produced and detected through
charged-current (CC) weak-interaction processes
described by the CC weak-interaction Lagrangian
\begin{align}
\mathcal{L}_{\text{CC}}(x)
=
\null & \null
\frac{g}{\sqrt{2}}
\sum_{a=e,\mu}
\overline{a_{L}}(x) \gamma^{\rho} \nu_{aL}(x) W_{\rho}(x)
+
\text{H.c.}
\nonumber
\\
=
\null & \null
\frac{g}{\sqrt{2}}
\sum_{a=e,\mu}
\sum_{k=1,2}
\overline{a_{L}}(x) \gamma^{\rho} U_{ak} \nu_{kL}(x) W_{\rho}(x)
+
\text{H.c.}
\,,
\label{002}
\end{align}
where $g$ is the coupling constant.
In general, the $2\times2$ unitary mixing matrix can be written as
\begin{equation}
U
=
\begin{pmatrix}
\cos\vartheta e^{i\omega_{1}} & \sin\vartheta e^{i(\omega_{1}+\alpha)}
\\
- \sin\vartheta e^{i(\omega_{2}-\alpha)} & \cos\vartheta e^{i\omega_{2}}
\end{pmatrix}
\,.
\label{003}
\end{equation}
However,
two of the three phases are unphysical,
because they can be eliminated by rephasing the two Dirac charged-lepton fields
without affecting the respective kinetic and mass Lagrangians
(the free Dirac Lagrangian is invariant under rephasing of the field)
and the Lagrangians of the other interactions to which charged leptons take part
(electromagnetic and weak neutral-current).
On the other hand,
a Majorana field cannot be rephased,
because the Majorana mass term is not invariant under rephasing of the field
(see Ref.~\cite{Giunti-Kim-2007}).
Therefore,
if the massive neutrino fields are of Majorana type,
one of the three phases in Eq.~(\ref{003}),
or a linear combination of them,
is physical \cite{Bilenky:1980cx,Doi:1980yb,Schechter:1980gr,Schechter:1981gk}.
This phase is called ``Majorana phase''.
The choice of the Majorana phase is arbitrary.
Let us consider two possibilities:

\begin{enumerate}

\renewcommand{\labelenumi}{(\theenumi)}
\renewcommand{\theenumi}{\arabic{enumi}}

\item
\label{004}
Rephasing the charged-lepton fields by
\begin{equation}
e_{L}(x) \to e^{i\omega_{1}} e_{L}(x)
\,,
\qquad
\mu_{L}(x) \to e^{i(\omega_{2}-\alpha)} \mu_{L}(x)
\,,
\label{005}
\end{equation}
we obtain
\begin{equation}
U^{(\text{\ref{004}})}
=
\begin{pmatrix}
\cos\vartheta & \sin\vartheta e^{i\alpha}
\\
- \sin\vartheta & \cos\vartheta e^{i\alpha}
\end{pmatrix}
\,.
\label{006}
\end{equation}
This is the most convenient choice,
because it allows to write the mixing matrix as
\begin{equation}
U^{(\text{\ref{004}})}
=
\begin{pmatrix}
\cos\vartheta & \sin\vartheta
\\
- \sin\vartheta & \cos\vartheta
\end{pmatrix}
\begin{pmatrix}
1 & 0
\\
0 & e^{i\alpha}
\end{pmatrix}
=
R(\vartheta) D(\alpha)
\,.
\label{007}
\end{equation}
In this form,
the Majorana phase $\alpha$ is factorized in the diagonal matrix
$ D(\alpha) = \text{diag}(1,e^{i\alpha}) $
on the right of the mixing matrix.

\item
\label{008}
Rephasing the charged-lepton fields by
\begin{equation}
e_{L}(x) \to e^{i\omega_{1}} e_{L}(x)
\,,
\qquad
\mu_{L}(x) \to e^{i\omega_{2}} \mu_{L}(x)
\,,
\label{009}
\end{equation}
we obtain
\begin{equation}
U^{(\text{\ref{008}})}
=
\begin{pmatrix}
\cos\vartheta & \sin\vartheta e^{i\alpha}
\\
- \sin\vartheta e^{-i\alpha} & \cos\vartheta
\end{pmatrix}
=
D^{\dagger}(\alpha) R(\vartheta) D(\alpha)
\,.
\label{010}
\end{equation}
This is the choice adopted in Ref.~\cite{0912.5266}.

\end{enumerate}

In order to investigate if there is any effect due to the Majorana phase $\alpha$
in neutrino oscillations,
following Ref.~\cite{0912.5266}
we consider an initial generic neutrino state which is a superposition of
$\nu_{e}$ and $\nu_{\mu}$:
\begin{equation}
| \nu(0) \rangle
=
A_{e} | \nu_{e} \rangle
+
A_{\mu} | \nu_{\mu} \rangle
\,.
\label{011a}
\end{equation}
Since the mixing of states is given by
\begin{equation}
| \nu_{a} \rangle
=
\sum_{k=1,2} U_{ak}^{*} | \nu_{k} \rangle
\qquad
(a=e,\mu)
\,,
\label{011b}
\end{equation}
we have
\begin{equation}
| \nu(0) \rangle
=
\left( A_{e} U_{e1}^{*} + A_{\mu} U_{\mu1}^{*} \right) | \nu_{1} \rangle
+
\left( A_{e} U_{e2}^{*} + A_{\mu} U_{\mu2}^{*} \right) | \nu_{2} \rangle
\,.
\label{011c}
\end{equation}
The probability of $\nu_{e}$ detection at a time $t$ is given by
\begin{equation}
P_{\nu_{e}}(t)
=
| \langle \nu_{e} | \nu(t) \rangle |^2
\,.
\label{012a}
\end{equation}
Considering ultrarelativistic neutrinos, we obtain
\begin{equation}
P_{\nu_{e}}(t)
=
\left|
A_{e} |U_{e1}|^2 + A_{\mu} U_{e1} U_{\mu1}^{*}
+
\left( A_{e} |U_{e2}|^2 + A_{\mu} U_{e2} U_{\mu2}^{*} \right)
\exp\!\left( - i \frac{ \Delta{m}^2 t }{ 2 E } \right)
\right|^2
\,,
\label{012b}
\end{equation}
where
$ \Delta{m}^{2} = m_{2}^2 - m_{1}^2 $
and
$E$ is the neutrino energy.

Using the choice of phases $U^{(\text{\ref{004}})}$
of the mixing matrix,
it is clear that there is no effect of the Majorana phase in the oscillation probability,
since the contribution of $\alpha$ in $U_{e2}$ and $U_{\mu2}^{*}$ cancels in the product:
\begin{equation}
P_{\nu_{e}}^{(\text{\ref{004}})}(t)
=
\left|
A_{e} \cos^2\vartheta - A_{\mu} \cos\vartheta \sin\vartheta
+
\left( A_{e} \sin^2\vartheta + A_{\mu} \cos\vartheta \sin\vartheta \right)
\exp\!\left( - i \frac{ \Delta{m}^2 t }{ 2 E } \right)
\right|^2
\,.
\label{013}
\end{equation}

This is enough to confute the argument presented in
Ref.~\cite{0912.5266}
that it is possible to observe the effect of the Majorana phases of the neutrino mixing matrix
in a neutrino oscillation experiment with an initial beam described by a superposition
of different neutrino flavors.
One can see it also by noting that
the probability $P_{\nu_{e}}(t)$ in Eq.~(\ref{012b}) is invariant under the rephasing
\begin{equation}
U_{ak} \to e^{i\phi_{k}} U_{ak}
\qquad
(a=e,\mu;\,k=1,2)
\,,
\label{028}
\end{equation}
which corresponds to the freedom to rephase arbitrarily the massive neutrino fields in the Dirac case,
when the Majorana phase is unphysical.
Indeed, a rephasing of this type,
$ U_{a2}^{(\text{\ref{004}})} \to e^{-i\alpha} U_{a2}^{(\text{\ref{004}})} $
for $a=e,\mu$,
eliminates the Majorana phase in Eq.~(\ref{006}).

It is however interesting to understand why the authors of Ref.~\cite{0912.5266}
have been led to an incorrect conclusion
as a consequence of their choice of phases $U^{(\text{\ref{008}})}$
of the mixing matrix.
With this choice of phases the contribution of $\alpha$ does not cancel in Eq.~(\ref{012b}):
\begin{align}
P_{\nu_{e}}^{(\text{\ref{008}})}(t)
=
\null & \null
\Big|
A_{e} \cos^2\vartheta - A_{\mu} e^{i\alpha} \cos\vartheta \sin\vartheta
\nonumber
\\
\null & \null
+
\left( A_{e} \sin^2\vartheta + A_{\mu} e^{i\alpha} \cos\vartheta \sin\vartheta \right)
\exp\!\left( - i \frac{ \Delta{m}^2 t }{ 2 E } \right)
\Big|^2
\,.
\label{014}
\end{align}

The difference between $P_{\nu_{e}}^{(\text{\ref{004}})}(t)$ and $P_{\nu_{e}}^{(\text{\ref{008}})}(t)$
is an apparent paradox,
because different choices of the phases
of the mixing matrix cannot induce different physical effects.

A hint towards the solution of the paradox comes from the fact that in Eq.~(\ref{014})
the Majorana phase appears always in the product
$ A_{\mu} e^{i\alpha} $,
i.e. as a change of phase of $A_{\mu}$.
Hence one is led to ask if the phases of the coefficients $A_{e}$ and $A_{\mu}$ in the initial neutrino state
in Eq.~(\ref{011a})
have physical meaning.
The answer is that these phases have no physical meaning,
because of the phase freedom of the charged-lepton fields,
which induces a phase freedom in the corresponding charged-lepton states and in the associated
flavor neutrino state.
In order to clarify this statement,
let us consider the complete leptonic state produced in a CC interaction process,
for example in $W^{+}$ decay:
\begin{equation}
| \psi(0) \rangle
=
A_{e} | e^{+}, \nu_{e} \rangle
+
A_{\mu} | \mu^{+}, \nu_{\mu} \rangle
\,,
\label{015}
\end{equation}
with
\begin{equation}
A_{a}
\propto
\langle a^{+}, \nu_{a} | \overline{\nu_{aL}}(0) \gamma^{\rho} a_{L}(0) W_{\rho}^{\dagger}(0) | W^{+} \rangle
\qquad
(a=e,\mu)
\,,
\label{016}
\end{equation}
at first order in perturbation theory,
neglecting the effects of the neutrino masses in the decay process.
From Eq.~(\ref{016}) one can see that the phase freedom of the charged-lepton fields
implies that the phases of the initial amplitudes
$A_{e}$
and
$A_{\mu}$
are arbitrary and unphysical.

This means that the state (\ref{011a}) cannot be realized in nature.
In other words, it is not possible to create a coherent superposition
of
$ | \nu_{e} \rangle $
and
$ | \nu_{\mu} \rangle $.
In fact,
if the charged lepton is detected,
the neutrino state is reduced to either
$ | \nu_{e} \rangle $
and
$ | \nu_{\mu} \rangle $.
On the other hand, if the charged lepton is not detected,
one must sum all neutrino interaction probabilities over the different charged lepton contributions.
This is equivalent of considering the neutrino state as an incoherent superposition of
$ | \nu_{e} \rangle $
and
$ | \nu_{\mu} \rangle $,
which is better described by a density matrix.
In this case we have
\begin{equation}
P_{\nu_{e}}(t)
=
\sum_{a=e,\mu}
\left|
\langle a^{+}, \nu_{e} | \psi(t) \rangle
\right|^2
\,,
\label{017}
\end{equation}
with $ | \psi(t) \rangle $ given by,
from Eq.~(\ref{015}),
\begin{align}
| \psi(t) \rangle
=
\null & \null
\left(
A_{e} U_{e1}^{*} | e^{+}, \nu_{1} \rangle
+
A_{\mu} U_{\mu1}^{*} | \mu^{+}, \nu_{1} \rangle
\right)
e^{ - i E_{1} t }
\nonumber
\\
\null & \null
+ \left(
A_{e} U_{e2}^{*} | e^{+}, \nu_{2} \rangle
+
A_{\mu} U_{\mu2}^{*} | \mu^{+}, \nu_{2} \rangle
\right)
e^{ - i E_{2} t }
\,,
\label{018}
\end{align}
where $E_{1}$ and $E_{2}$ are, respectively, the energies of $\nu_{1}$ and $\nu_{2}$,
such that for the ultrarelativistic neutrinos under consideration
$ E_{2} - E_{1} \simeq \Delta{m}^2 / 2 E $.
The amplitudes in Eq.~(\ref{017}) are given by
\begin{align}
\null & \null
\langle e^{+}, \nu_{e} | \psi(t) \rangle
=
A_{e}
\left(
|U_{e1}|^2 e^{ - i E_{1} t }
+
|U_{e2}|^2 e^{ - i E_{2} t }
\right)
\,,
\label{019}
\\
\null & \null
\langle \mu^{+}, \nu_{e} | \psi(t) \rangle
=
A_{\mu}
\left(
U_{e1} U_{\mu1}^{*} e^{ - i E_{1} t }
+
U_{e2} U_{\mu2}^{*} e^{ - i E_{2} t }
\right)
\,.
\label{020}
\end{align}
Inserting these two expressions in Eq.~(\ref{017}), we obtain
\begin{equation}
P_{\nu_{e}}(t)
=
|A_{e}|^2 P_{\nu_{e}\to\nu_{e}}(t)
+
|A_{\mu}|^2 P_{\nu_{\mu}\to\nu_{e}}(t)
\,,
\label{021}
\end{equation}
with the standard transition probabilities
\begin{equation}
P_{\nu_{e}\to\nu_{e}}(t) = 1 - \sin^2 2\vartheta \sin^2\!\left( \frac{ \Delta{m}^2 t }{ 4 E } \right)
\,,
\qquad
P_{\nu_{\mu}\to\nu_{e}}(t) = \sin^2 2\vartheta \sin^2\!\left( \frac{ \Delta{m}^2 t }{ 4 E } \right)
\,.
\label{022}
\end{equation}
These probabilities
are independent of the Majorana phase $\alpha$
in both the choices of phases
$U^{(\text{\ref{004}})}$
and
$U^{(\text{\ref{008}})}$
of the mixing matrix,
as well as in any other choice allowed by the rephasing freedom of the charged-lepton fields.
Hence we see that the correct probability to detect a $\nu_{e}$
is independent of the Majorana phase and is independent of the phases of the initial amplitudes
$A_{e}$
and
$A_{\mu}$.
Thus,
the initial neutrino created in a charged-current interaction process which can produce both
$\nu_{e}$ and $\nu_{\mu}$
is not described correctly by the coherent state in Eq.~(\ref{011a}),
but can be described by the density matrix operator
\begin{equation}
\hat\rho(0)
=
|A_{e}|^2 | \nu_{e} \rangle \langle \nu_{e} |
+
|A_{\mu}|^2 | \nu_{\mu} \rangle \langle \nu_{\mu} |
\,.
\label{023}
\end{equation}

Another way of seeing that the probability $P_{\nu_{e}}(t)$ in Eq.~(\ref{012b}) is unobservable
is to notice that it is not invariant under the rephasing
\begin{equation}
U_{ak} \to e^{i\phi_{a}} U_{ak}
\qquad
(a=e,\mu;\,k=1,2)
\,,
\label{028a}
\end{equation}
which corresponds to the freedom to rephase arbitrarily the charged-lepton fields,
as done, for example, in Eqs.~(\ref{005}) and (\ref{009}).
In fact, this lack of invariance is the origin of the difference between
$P_{\nu_{e}}^{(\text{\ref{004}})}(t)$
in Eq.~(\ref{013})
and
$P_{\nu_{e}}^{(\text{\ref{008}})}(t)$
in Eq.~(\ref{014}).

Let us finally investigate if the Majorana phase has a physical effect
in oscillations of a beam of neutrinos
produced in a neutral-current interaction process,
which is the only possibility of neutrino production (and detection)
besides the charged-current processes considered above.
In a neutral-current process,
as for example $Z^{0}$ decay,
neutrinos and antineutrinos are produced together and the different mass or flavor eigenstates
have the same initial amplitude,
because neutral current interactions are flavor-blind.
In the two-neutrino framework the initial neutrino state is
\begin{equation}
| \psi(0) \rangle
=
\frac{1}{\sqrt{2}}
\left(
| \nu_{1}, \bar\nu_{1} \rangle
+
| \nu_{2}, \bar\nu_{2} \rangle
\right)
=
\frac{1}{\sqrt{2}}
\left(
| \nu_{e}, \bar\nu_{e} \rangle
+
| \nu_{\mu}, \bar\nu_{\mu} \rangle
\right)
\,.
\label{024}
\end{equation}
It was shown in Ref.~\cite{Smirnov:1991eg} that flavor transitions
occur only if both the neutrino and the antineutrino are detected.
Hence, we consider the case of detection of a $\nu_{a}$ at the time $t$
and a $\bar\nu_{b}$ at the time $\bar{t}$,
with $a,b=e,\mu$.
The relevant neutrino state is
\begin{equation}
| \psi(t,\bar{t}) \rangle
=
\frac{1}{\sqrt{2}}
\left(
| \nu_{1}, \bar\nu_{1} \rangle
e^{ - i ( E_{1} t + \bar{E}_1 \bar{t}) }
+
| \nu_{2}, \bar\nu_{2} \rangle
e^{ - i ( E_{2} t + \bar{E}_2 \bar{t}) }
\right)
\,,
\label{025}
\end{equation}
where
$\bar{E}_{1}$ and $\bar{E}_{2}$ are, respectively, the energies of $\bar\nu_{1}$ and $\bar\nu_{2}$,
such that for the ultrarelativistic antineutrinos under consideration
$ \bar{E}_{2} - \bar{E}_{1} \simeq \Delta{m}^2 / 2 \bar{E} $.
The detection probability is given by
\begin{align}
P_{ \nu_{a}, \bar\nu_{b} }(t,\bar{t})
=
\null & \null
| \langle \nu_{a}, \bar\nu_{b} | \psi(t,\bar{t}) \rangle |^2
\nonumber
\\
=
\null & \null
\frac{1}{2}
\left(
|U_{a1}|^2 |U_{b1}|^2
+
|U_{a2}|^2 |U_{b2}|^2
\right)
\nonumber
\\
\null & \null
+
\text{Re}\left\{
U_{a1} U_{b1}^{*} U_{a2}^{*} U_{b2}
\exp\!\left[ - i \left( \frac{ \Delta{m}^2 t }{ 2 E } + \frac{ \Delta{m}^2 \bar{t} }{ 2 \bar{E} } \right) \right]
\right\}
\,.
\label{026}
\end{align}
We confirm that flavor transitions
occur only if both the neutrino and the antineutrino are detected,
as shown in Ref.~\cite{Smirnov:1991eg},
because when the neutrino or antineutrino are not detected, the corresponding probabilities are
\begin{equation}
P_{\bar\nu_{b}}
=
\sum_{a=e,\mu} P_{ \nu_{a}, \bar\nu_{b} }(t,\bar{t})
=
\frac{1}{2}
\,,
\qquad
P_{\nu_{a}}
=
\sum_{b=e,\mu} P_{ \nu_{a}, \bar\nu_{b} }(t,\bar{t})
=
\frac{1}{2}
\,.
\label{027}
\end{equation}
These are the same flavor probabilities as those in the initial state in Eq.~(\ref{024}).

Moreover,
the probability $P_{ \nu_{a}, \bar\nu_{b} }(t,\bar{t})$ is invariant under the rephasing in Eq.~(\ref{028}).
Therefore, it is independent of the Majorana phase.
One can also verify it explicitly using the two choices of phases
$U^{(\text{\ref{004}})}$
and
$U^{(\text{\ref{008}})}$
of the mixing matrix.

In conclusion,
we have shown that the Majorana phases of the neutrino mixing matrix
cannot have any effect in neutrino oscillations,
in agreement with Refs.~\cite{Bilenky:1980cx,Doi:1980yb,Schechter:1981gk,Langacker:1986jv}
and
contrary to the argument recently presented in Ref.~\cite{0912.5266}.
We have also shown that
in a charged-current weak interaction process it is not possible to
create a coherent superposition of different flavor neutrinos which is independent of the associated
charged leptons.

\raggedright

\end{document}